\documentclass[jou,floatsintext,noextraspace,a4paper]{apa6}
\usepackage{natbib}
\usepackage{standalone}
\usepackage{times}
\usepackage[utf8]{inputenc}
\usepackage{textcomp}
\usepackage{lineno}
\usepackage{color,soul}
\usepackage[]{graphicx}
\usepackage{standalone}
\usepackage{graphicx}
\title{Statistical reform and the replication crisis}
\shorttitle{Statistical reform}
\twoauthors{Lincoln J Colling$^\star$}{Dénes Sz\H{u}cs}
\leftheader{Colling, and Sz\H{u}cs}
\twoaffiliations{Department of Psychology\\University of Cambridge}{Department of Psychology\\University of Cambridge}
\affiliation{University of Cambridge}
\abstract{The replication crisis has prompted many to call for statistical reform within the psychological sciences. Here we examine issues within Frequentist statistics that may have led to the replication crisis, and we examine the alternative---Bayesian statistics---that many have suggested as a replacement. The Frequentist approach and the Bayesian approach offer radically different perspectives on evidence and inference with the Frequentist approach prioritising error control and the Bayesian approach offering a formal method for quantifying the relative strength of evidence for hypotheses. We suggest that rather than mere statistical reform, what is needed is a better understanding of the different modes of statistical inference and a better understanding of how statistical inference relates to scientific inference.}
\journal{To appear in Review of Philosophy and Psychology}

\authornote{$^\star$Correspondence concerning this article should be addressed to Lincoln J Colling, Downing Street, CB2 3EB, Cambridge, UK.\\ E-mail: lincoln@colling.net.nz}
\begin{document}
\maketitle

\section{Introduction}\label{introduction}

A series of events in the early 2010s, including the publication of Bem's \citeyearpar{Bem:2011jz} infamous study on extrasensory perception (or PSI), and data fabrication by Diederik Stapel and others \citep{Stroebe:2012hi}, led some prominent researchers to claim that psychological science was suffering a ``crisis of confidence'' \citep{Pashler:2012eq}. At the same time as these scandals broke, a collective of scientists was formed to undertake a large-scale collaborative attempt to replicate findings published in three prominent psychology journals \citep{OpenScienceCollaboration:2012bt}. The results of these efforts would strike a further blow to confidence in the field \citep{Yong:2012ja}, and with the replication crisis in full swing old ideas that science was self-correcting seemed to be on shaky ground \citep{Ioannidis:2012em}.

One of the most commonly cited causes of the replication crisis has been the statistical methods used by scientists, and this has resulted in calls for statistical reform \citep[e.g.,][]{Wagenmakers:2011ej, Haig:2016kh, Dienes:2011cd}. Specifically, the suite of procedures known as Null Hypothesis Significance Testing (NHST), or simply \emph{significance testing}, and their associated \emph{p} values, and claims of statistical significance, have come in most to blame \citep{Nuzzo:2014bp}. The controversy surrounding significance testing and \emph{p} values is not new \citep[see][ for a detailed treatment]{Nickerson:2000jl}; however, the replication crisis has resulted in renewed interest in the conceptual foundations of significance testing and renewed criticism of the procedures themselves \citep[e.g.,][]{Wagenmakers:2007bg, Dienes:2011cd, Szucs:2017it}. Some journals have gone so far as to ban \emph{p} values from their pages \citep{Trafimow:2014bo} while others have suggested that what gets to be called \emph{statistically significant} should be redefined \citep{Benjamin:2017gh}. Some criticism of \emph{p} values stems from the nature of \emph{p} values themselves---a position particularly common with those advocating some form of Bayesian statistics---while other criticisms have focused on their use rather than attacking the conceptual grounding of the procedures themselves \citep{Nickerson:2000jl, GarciaPerez:2016cr}. However, one thing that was made clear by the replication crisis, and the ensuing debates about the use of \emph{p} values, is that few people understood the nature of \emph{p} values, the basis of the Frequentist statistics that generate them, and what inferences could be warranted on the basis of \emph{statistical significance}. Such was the confusion and misunderstanding among many in the scientific community that the American Statistical Association (ASA) took the unusual step of releasing a statement on statistical significance and \emph{p} values in the hope of providing some clarity about their meaning and use \citep{Wasserstein:2016joa}.

In order to make sense of the criticisms of \emph{p} values and to make sense of their role in the replication crisis it is important to understand what a \emph{p} value is (how it is derived) and what conditions underwrite its inferential warrant. We detail this in Section 2. There we also outline what inferences can be made on the basis of \emph{p} values and introduce a recent framework, the \emph{error statistical approach}, which addresses some of the shortcomings of previous Frequentist approaches. In Section 3 we introduce an alternative to Frequentist statistics---Bayesian statistics. Specifically, in Section 3.1 we examine some of the claimed benefits of the Bayesian approach while in Section 3.2 we introduce the Bayesian notion of statistical evidence, and examine whether the Bayesian approach and the Frequentist approach lead to different conclusions. In Section 4 we compare the two approaches more directly and examine how each approach fits into a system of scientific inference. Finally, we conclude by suggesting that rather than mere \emph{statistical reform} what is needed is a change in how we make scientific inferences from data. And we suggest that there might be benefits in pragmatic pluralism in statistical inference.

\section{Frequentist statistic and \emph{p} values}{Frequentist statistic and p values}\label{frequentist-statistic-and-p-values}

The ASA statement on \emph{p} values provides an informal definition of a \emph{p} value as ``the probability \emph{under a specified statistical model} that a statistical summary of the data (e.g., the sample mean difference between two compared groups) would be \emph{equal to or more extreme} than its observed value'' \citep[our emphasis]{Wasserstein:2016joa}. Probability is an ill-defined concept with no generally agreed definition that meets all the requirements that one would want. In the context of significance testing, however, \emph{p} values are often interpreted with reference to the long run behaviour of the test procedure \citep[e.g., see][]{Neyman:1933ff}. That is, they can be given a \emph{frequency} interpretation \citep[see][ for more detail on a frequency interpretation of confidence intervals]{Morey:2016ca}. Although a frequency interpretation may not be universally accepted (or acknowledged), this interpretation more clearly highlights the link between \emph{p} values and the long run behaviour of significance tests. When given a frequency interpretation, the \emph{p} indicates how often under a specified model, considering repeated experiments, a test statistic as large or larger than the one observed would be observed if it was the case that the null hypothesis (for example, the hypothesis that the two groups are drawn from the same population) was true. The \emph{p} value is calculated from the \emph{sampling distribution}, which describes what is to be expected \emph{over the long run} when samples are tested.

What allows one to draw inferences from \emph{p} values is the fact that statistical tests should rarely produce small \emph{p} values if the null model is true, and provided certain conditions are met\footnote{These conditions are the \emph{assumptions} of the statistical test. These might include things such as equal variance between the two groups in the case of \emph{t} tests or certain assumptions about the covariance matrix in the case of factorial ANOVA. These are often violated and, therefore, tests can be inaccurate. Correction procedures, tests that are robust to violations, or tests that generate their own sampling distribution from the data (such as randomisation tests) are available. However, we will not discuss these as our focus will primarily be on the \emph{inferences} that statistical tests support.}. It is also this fact that leads to confusion. Specifically, it leads to the confusion that if a small \emph{p} is obtained then one can be 1 - \emph{p} sure that the alternative hypothesis is true. This common misunderstanding can result in an interpretation that, for example, \emph{p} = 0.01 indicates a 99\% probability that the detected effect is real. However, to conclude this would be to confuse the probability of obtaining the data (or more extreme) given the null hypothesis with the probability that the null hypothesis is true given the data \citep[see][ for examples of this confusion]{Nickerson:2000jl}.

The confusion that \emph{p} values warrant inferences they do not has similarly led to confusion about the conditions under which \emph{p} values \emph{do} warrant inferences. We will explain what inferences \emph{p} values do warrant in Section 2.3, but before this can be done it is important to understand what conditions must be met before they can support \emph{any} inferences at all. For now, however, it is sufficient to know that inferences on the basis of \emph{p} values rely on the notion of \emph{error control}. As we will see, violations of the conditions that grant these error control properties may be common.

\subsection{Controlling false positives}\label{controlling-false-positives}

The first condition under which \emph{p} values are able to provide information on which to base inferences is that \emph{if} the null hypothesis is true then \emph{p} values should be uniformly distributed\footnote{We should note that this is only generally true when the null model takes the form of a continuous probability distribution, which is common for the statistical procedures used in psychology. This assumption does not necessarily hold for discrete probability distributions.}. For instance, if one was to repeatedly draw samples from a standard normal distribution centred on 0, and after each sample test the null hypothesis that \(\mu = 0\) (for example, by using a one sample \emph{t}-test) one would obtain a distribution of \emph{p} values approximately like the one shown in Figure 1 (A). This fact appears to contradict at least one common misinterpretation of \emph{p} values, specifically the expectation that routinely obtaining high \emph{p} values should be common when the null hypothesis is true---for instance the belief that obtaining \emph{p} \textgreater{} .90 should be common when the null is true and \emph{p} \textless{} .10 should be rare, when in fact they will occur with equal frequency \citep[see][ for common misinterpretations of \emph{p} values]{Nickerson:2000jl}. Herein lies the concept of the \emph{significance threshold}. While, for instance, \emph{p} \(\approx\) .87, and \emph{p} \(\approx\) .02 will occur with equal frequency if the null is true, \emph{p} values less than the threshold (defined as \(\alpha\)) will only occur with the frequency defined by that threshold. Provided this condition is met, this sets an \emph{upper bound} on how often one will incorrectly infer the presence of an effect when in fact the null is true.

\begin{figure*}
\centering
\includegraphics[width=.95\textwidth]{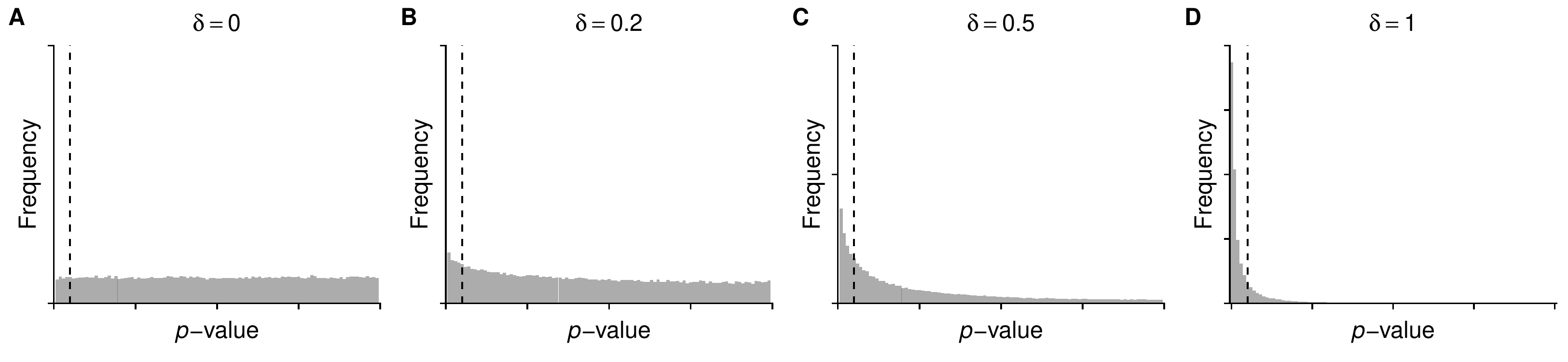}
\caption{Examples of \emph{p} value distributions under different effect sizes. An effect size of \(\delta=0\) indicates that the null hypothesis is true.}
\end{figure*}

The uniformity of the \emph{p} value distribution under the null hypothesis is, however, only an ideal. In reality, there are many behaviours that researchers can engage in that can change this distribution. These behaviours, which have been labelled \emph{p} hacking, QPRs (questionable research practices), data dredging, and significance chasing, therefore threaten to revoke the \emph{p} value's inferential licence\footnote{Concerns over these behaviours is not new. Referring to the practice as ``cooking'', Charles Babbage \citeyearpar{Babbage:1830uu} noted that one must be very unlucky if one is unable to select only agreeable observations out of the multitude that one has collected.} \citep[e.g.,][]{Ware:2015fn, Szucs:2016cc, Simmons:2011iw}. One of the most common behaviours is optional stopping (also known as data peaking). To illustrate this behaviour, we will introduce an example, which we will return to later in the context of Bayesian alternatives to significance testing. Consider Alice who collects a sample of 10 observations. After collecting her sample, she conducts a significance test to determine whether the mean is significantly different from some null value (this need not be zero, but often this is the case). Upon finding \emph{p} = .10, she decides to add more observations checking after adding each additional observation whether \emph{p} \textless{} .05. Eventually, this occurs after she has collected a sample of 20.

On a misunderstanding of the \emph{p} value this behaviour seems innocuous, so much so that people often express surprise when they are told it is forbidden \citep[e.g.,][]{John:2012eo, Yu:2013dx}. However, it only seems innocuous on the incorrect assumption that large \emph{p} values should be common if the null is true. After all, Alice checked her \emph{p} values after every few samples, and while they may have changed as each new sample was added, they weren't \emph{routinely} large. However, optional stopping distorts the distribution of \emph{p}+values so that it is no longer uniform. Specifically, the probability of obtaining \emph{p} \textless{} \(\alpha\), when optional stopping is applied, is no longer equal to \(\alpha\) and instead it can be dramatically higher than \(\alpha\)\footnote{To illustrate this, we conducted a simple simulation. We drew samples (\emph{n} = 1000) from a standard normal distribution centred at zero. The values were then tested, using a one sample \emph{t}-test against the null hypothesis that \(\mu = 0\) by first testing the first 10 values, then the first 11, the first 12 and so forth until either obtaining a \emph{p} \textgreater{} 0.05 or exhausting the 1000 samples. After repeating this procedure 10000 times, we were able to obtain a significant \emph{p} value approximately 46\% of the time. The median sample size for a significant result was 56.}. Thus, in the case of optional stopping, the connection between the value of \emph{p} and the frequency of obtaining a \emph{p} value of that magnitude (or smaller) is broken.

A related issue that can revoke the inferential licence of \emph{p} values occurs when a researcher treats a collection of \emph{p} values (also known as a family of tests) in the same way they might treat a single \emph{p} value. Consider the case where a researcher runs ten independent statistical tests. Given the null, the frequency of finding a significant result (\emph{p} \textless{} 0.05) is 5\% \emph{for each test}. As a result, the chance of finding \emph{at least one} significant effect in a family of 10 tests is approximately 40\%. While most researchers understand the problem of confusing the \emph{chance of finding a significant test} with the \emph{chance of finding at least one significant test in a collection of tests}, in the context of simple tests like \emph{t}-tests, this confusion persists in more complex situations like factorial ANOVA. Consider a two factor ANOVA, which produces three test statistics: Researchers can make this error and confuse the chance of finding \emph{at least one} significant test (for example, a main effect or interaction) with the chance of \emph{a particular} test being significant. In the case of the former, the chance of finding at least one significant main effect or interaction in a two factor ANOVA can be approximately 14\%. That a recent survey of the literature, which accompanied a paper pointing out this hidden multiplicity, found that only around 1\% of researchers (across 819 papers in six leading psychology journals) took this into account when interpreting their data demonstrates how widespread this confusion is \citep{Cramer:2015jv}. Furthermore, high profile researchers have expressed surprise upon finding this out \citep{Bishop:2014ee}, further suggesting that it was not commonly known. As noted by \citet{Bishop:2014ee}, this problem might be particularly acute in fields like event-related potential (ERP) research where researchers regularly analyse their data using large factorial ANOVAs and then interpret whatever results fall out. As many as four factors are not uncommon, and consequently, the chance of finding at least one significant effect can be roughly the same as correctly calling a coin flip. Furthermore, if a theory can be supported by a main effect of one factor, or any interaction involving that factor---that is, if one \emph{substantive hypothesis} can be supported by multiple \emph{statistical hypotheses}---then in the case of a four-way ANOVA that theory will find support as often as 25\% of the time even if the null hypothesis is true.

With this in mind, the advice offered by \citet{Bem:2009bx} appears particularly unwise: In speaking about data that might have been collected from an experiment, he suggests ``{[}e{]}xamine them from every angle. Analyze the sexes separately. Make up new composite indices.'' (pp.~4--5). That is, add additional factors to the ANOVA to see if anything pops up. However, as we have seen, adding additional factors simply increases the chance of significance even when the null is true. This hidden multiplicity is rarely acknowledged in scientific papers. More generally, any \emph{data dependent} decisions---for example, choosing one composite index over another based on the data---greatly increases the chance of finding significance regardless of whether multiple comparisons \emph{were actually made}.\footnote{In addition to specific data dependent decisions, \citet{Steegen:2016jy} outline how a number of seemingly arbitrary decisions made during the analysis process can give rise to a very wide range of results.} Indeed, \citet[p 6]{Bem:2009bx} goes on to state that:

\begin{quote}
``Scientific integrity does not require you to lead your readers through all your wrongheaded hunches only to show---voila!---they were wrongheaded. A journal article should not be a personal history of your stillborn thoughts.''
\end{quote}

While such a journal article may make for tedious reading, it is only by including all those thoughts, those wrongheaded hunches, those data dependent decisions, that will allow the reader to determine whether the process by which the results were obtained deserve to be awarded any credibility, or whether they are as impressive as correctly calling a coin flip.

\subsection{Controlling false negatives}\label{controlling-false-negatives}

A second condition that must be met for inferences on the basis of \emph{p} values to be warranted is that low \emph{p} values (i.e., \emph{p} \(< \alpha\)) should occur more \emph{frequently} when a true or real effect is present. This occurs because when the discrepancy between the null model and the model from which the samples are actually drawn increases (something that can be quantified in terms of \emph{effect size}), the distribution of \emph{p} values, obtained in the long run, departs from uniformity. This is illustrated in Figure 1 (B--D) by showing the distribution of \emph{p} values obtained from repeated testing of samples drawn from distributions representing different true effect sizes. When a real effect is present, the frequency with which a \emph{p} value occurs is inversely proportional to its magnitude. This skewing of the \emph{p} value distribution in the presence of a real effect illustrates the concept of \emph{statistical power} \citep[e.g.,][]{Cohen:1992ev}. The greater the skew observed in the long run distribution of \emph{p} values the greater the statistical power. That is, power is equal to \(1 - \beta\), where \(\beta\) is the proportion of \emph{p} values \textgreater{} \(\alpha\) that occur when a true effect is present. Power, therefore, allows one to place an \emph{upper bound} on how often one will incorrectly conclude the \emph{absence} of an effect (of at least a particular magnitude) when in fact an effect (of that magnitude or greater) is present.

That \emph{p} values skew towards zero in the presence of a true effect implies that \emph{p} values near the threshold \(\alpha\) should be comparatively rare if a real effect is present. However, near threshold \emph{p} values are surprisingly common \citep{Masicampo:2012da, deWinter:2015bx}. This suggests that the reported effects may actually accord more with a true null hypothesis. However, they may also imply that statistical power is very low and that the distribution of \emph{p} values has not departed sufficiently from uniformity. Adequate statistical power---that is, the requirement that experiments are so designed such that in the long run they will produce an extremely skewed distribution of \emph{p} values---is a fundamental requirement if inferences are to be drawn on the basis of \emph{p} values. However, empirical studies of the scientific literature suggest that this requirement is not routinely met. For example, studies by \citet{Button:2013dza} and \citet{Szucs:2017dz} suggest that studies with low statistical power are common in the literature. Recall, it is only when the two conditions are met---uniformly distributed \emph{p} values when the null is true and a heavily skewed \emph{p} value distribution when a real effect is present---that good inferences on the basis of \emph{p} values are possible. Neither of these conditions are commonly met and, therefore, the epistemic value of \emph{p} values is routinely undermined.

What is the cause of low statistical power? In our definition of power, we said that power was determined by the skew of the \emph{p} value distribution in the presence of a \emph{given} true effect. That is, if samples of a fixed size are repeatedly drawn and tested with a statistical test, and a true effect is present, how often \emph{p} \textless{} .05 occurs depends on the \emph{magnitude} of the true effect. To draw valid inferences from \emph{p} values, in the long run, one needs to know the magnitude of the effect that one is making inferences about. If the magnitude of the effect is small, then one needs more information (larger samples) to reliably detect its presence. When the magnitude of the effect is large, then you can generate reliable decisions using less information (smaller samples). However, it is important to note that basing effect size estimates for \emph{a priori} power analyses on published results can be very problematic because in the presence of publication bias (only publishing significant results) the published literature will invariably overestimate the real magnitude of the effect. That is, when power is low, statistical significance acts to select only those studies that report effect sizes larger than the true effect. Only through averaging together significant and non-significant effects can one get a good estimate of the actual effect size. Interestingly, an examination of replication attempts by \citet{Simonsohn:2015ic} suggests that in many cases, effect size estimates obtained from high-powered replications imply that the original studies reporting those effects were underpowered and, therefore, could not have reliably studied effects of those magnitudes.

\subsection{Frequentist inferences}\label{frequentist-inferences}

Inferences on the basis of \emph{p} values can be difficult and unintuitive. The problems that we've outlined above are not problems of significance testing per se, rather they are a result of the inferential heuristics that people apply when conducting experiments---heuristics such as, ``if it's nearly significant then collect more data'' or ``if I can obtain significance with a small sample then it's more likely that my hypothesis is true''. Part of the reason why people may employ inferential heuristics is that several distinct frameworks exist for drawing inferences on the basis of \emph{p} values and often these are not clearly distinguished in the statistics textbooks or statistics training. In some cases, researchers may even be unaware that different frameworks exist. The two most prominent frameworks are those of Fisher \citep[e.g.,][]{Fisher:1925ff} and Neyman and Pearson \citep[e.g.,][]{Neyman:1933ff}. Fisher's view of inference was simply that data must be given an opportunity to disprove (that is, reject or falsify) the null hypothesis (\(H_{0}\)). The innovation of Neyman and Pearson was to introduce the alternative hypothesis (\(H_{1}\)) and with it the concept of false alarms (\emph{errors of the first type}, or inferring the presence of an effect when the null hypothesis is true) and false negatives (\emph{errors of the second type}, or inferring the absence of an effect when the alternative hypothesis is true). They also saw a different role for the \emph{p} value. Fisher was concerned with the actual magnitude of the \emph{p} value. Neyman and Pearson, on the other hand, were concerned with whether the \emph{p} value crossed a threshold (\(\alpha\)). If the \emph{p} value was smaller than \(\alpha\) then one could reject \(H_{0}\) and if the \emph{p} value was greater than \(\alpha\) one could fail to reject \(H_{0}\)\footnote{\citet{Neyman:1933ff} use the terminology \emph{accept} \(H_{0}\). However, \citet{Neyman:1976rr} uses the terminology \emph{do not reject} \(H_{0}\). Furthermore, he goes on to state that his preferred terminology is \emph{no evidence against} \(H_{0}\). We follow \citet{Neyman:1976rr} in preferring the \emph{no evidence against} or \emph{do not reject} phrasing.}. By fixing \(\alpha\) and \(\beta\) (that is, by maximising statistical power) at particular levels they could fix the long run error control properties of statistical tests, resulting in rules that, if followed, would lead to inferences that would rarely be wrong. The type of inferences employed in practice, however, appear in many ways to be a hybrid of the two views \citep{Gigerenzer:1993hb}. A consequence of this is that many of the inferences drawn from significance tests have been muddled and inconsistent.

As a result, some have argued that significance tests need a clearer inferential grounding. One such suggestion has been put forward by Mayo \citep{Mayo:1996xx, Mayo:2006iu, Mayo:2011wm} in the form of her \emph{error-statistical} philosophy. As the name suggests, it builds on the insight of Neyman and Pearson that Frequentist inference relies on the long run error probabilities of statistical tests. In particular, it argues that for inferences on the basis of \emph{p} values to be valid (that is, have good long run performance) a researcher cannot simply draw inferences between a null (e.g., no difference) and an alternative which is simply its negation (e.g., a difference). Long run performance of significance tests can only be controlled when inferences are with reference to \emph{a specific alternative hypothesis}. And inferences about these specific alternatives are only well justified if they have passed \emph{severe} tests.

Mayo \citeyearpar{Mayo:2011wm} explains \emph{severity} informally with reference to a math test as a test of a student's math ability. The math test counts as a \emph{severe} test of a student's math ability if it is the case that obtaining a high score would be unlikely unless it was the case that the student actually had a high maths ability. Severity is thus a function of a specific test (the math test), a specific observation (the student's score), and a specific inference (that the student is very good at maths).

More formally, \citet{Mayo:2011wm} state the \emph{severity principle} as follows:

\begin{quote}
Data \(x_0\) (produced by process \(G\)) do not provide good evidence for the hypothesis \(H\) if \(x_0\) results from a test procedure with a very low probability or capacity of having uncovered the falsity of \(H\), even if \(H\) is incorrect.
\end{quote}

Or put positively:

\begin{quote}
Data \(x_0\) (produced by process \(G\)) provide good evidence for hypothesis \(H\) (just) to the extent that test \(T\) has severely passed \(H\) with \(x_0\).
\end{quote}

Severity is, therefore, a property of a specific test with respect to a specific inference (or hypothesis) and some data. It can be assessed qualitatively, as in the math test example above, or quantitatively through the sampling distribution of the test statistic. To illustrate how this works in practice, we can consider the following example \citep[adapted from][]{Mayo:2017hn}. A researcher is interested in knowing whether the IQ scores of some group are above average. According to the null model, IQ scores are normally distributed with a mean of 100 and a \emph{known} variance of \(15^{2}\). After collecting 100 scores (\emph{n} = 100), she tests the sample against the null hypothesis \(H_{0}:\mu = 100\) with the alternative hypothesis \(H_{1}:\mu > 100\). If the observed mean (\(\bar{x}\)) was 103, then a z-test would be significant at \(\alpha = .025\). From something like a classic Neyman-Pearson approach, the inference that would be warranted on the basis of this observation would be something like \emph{reject} \(H_{0}\) and conclude that the mean is greater than 100.

A \emph{severity} assessment, however, allows one to go further. Instead of merely concluding that the group's mean (\(\mu_{1}\)) is greater than the null value (\(\mu_{0}\)), one can instead use the observed result (\(\bar{x}\)) to assess specific alternate inferences about discrepancies (\(\gamma\)) from \(\mu_{0}\) of the form \(H_{1}:\mu > \mu_{0} + \gamma\). For example, one might want to use the observation (\(\bar{x} = 103\)) to assess the hypothesis \(H_{1}:\mu > \mu_{0} + 1\) or the hypothesis \(H_{1}:\mu > \mu_{0} + 3\). The severity associated with the inference \(\mu > 101\) would be 0.91\footnote{In the R statistics package, severity for a z-test can be calculated using the command, \texttt{pnorm(x.bar\ -\ (h0\ +\ gamma)/\ (sigma\ /\ sqrt(n)))}, where \texttt{x.bar} is the observed mean, \texttt{h0} is the null value, \texttt{sigma} is the population standard deviation, \texttt{n} is the sample size, and \texttt{gamma} is the deviation from the null value that one wishes to draw an inference about.}, while the severity associated with the inference that \(\mu > 103\) is 0.5. Thus, according to the severity principle, the observation that \(\bar{x} = 103\) provides us with better grounds to infer that \(\mu_{1}\) is at least 101 relative to an inference that it is at least 103.

\begin{figure*}
\centering
\includegraphics[width=.95\textwidth]{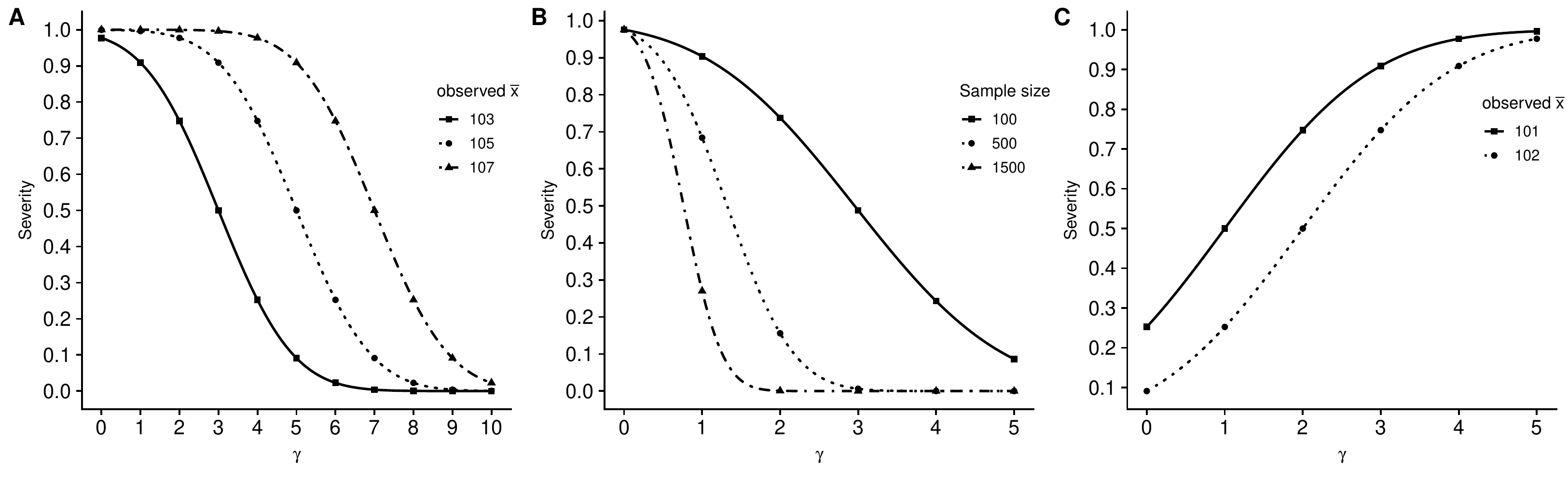}
\caption{Examples of severity curves for different statistically significant observations (A), barely significant observations with different sample sizes (B), and different non-significant observations (C).}
\end{figure*}

Just as one can use severity to test different inferences with respect to a fixed result, one can also use severity to assess a fixed inference with respect to different results. Consider again the inference that \(\mu > 103\). The severity associated with this inference and the result \(\bar{x} = 103\) is 0.5. However, if one had observed a different result of, for example, \(\bar{x} = 105\), then the severity associated with the inference \(\mu > 103\) would be 0.91. In order to visualise severity for a range of inferences with reference to a particular test and a particular observation, it is possible to plot severity as a function of the inference. Examples of different inferences about \(\mu\) for different observations (\(\bar{x}\)) is shown in Figure 2 (A).

The severity assessment of significant tests has a number of important properties. First, severity assessments license different inferences on the basis of different observed results. Consequently, rather than all statistically significant results being treated as equal, specific inferences may be more or less well justified on the basis of the specific \emph{p} value obtained. In our above example, the observation of \(\bar{x}=103\) (n = 100, \(\sigma=15\)) results in \emph{p} = .023, while the observation of \(\bar{x}=105\) results in \emph{p} \textless{} 0.001. Thus for a fixed n, lower \emph{p} values license inferences about larger discrepancies from the null. The severity assessment also highlights the distinction between statistical hypotheses and substantive scientific hypotheses. For example, a test of a scientific hypothesis might require that the data support inferences about some deviation from the null value that is at least of magnitude \(X\). The data might reach the threshold for statistical significance without the inference that \(\mu_{1} > \mu_{0} + X\) passing with high severity. Thus, the statistical hypothesis might find support without the theory being supported.

Severity assessments can also guard against unwarranted inferences in cases where the sample size is very large. Consider the case where one fixes the observed \emph{p} value (for example, to be just barely significant) and varies the sample size. What inferences can be drawn from these \emph{just} significant findings at these various sample sizes? On a simplistic account, all these significant tests warrant the inference \emph{reject} \(H_{0}\) and conclude some deviation (of unspecified magnitude) from the null. A severity assessment, however, allows for a more nuanced inference. As sample size increases, one would only be permitted to infer smaller and smaller discrepancies from the null with high severity. Again using our example above, the observation associated with \emph{p} = .025 and n = 100, allows one to infer that \(\mu_{1} > 101\) with a severity of 0.9. However, the same \emph{p} value obtained with n = 500 reduces the severity of the same inference to 0.68. An illustration of the influence of sample size on severity is shown in Figure 2 (B). If one wanted to keep the severity assessment high, one would need to change one's inference to, example, \(\mu > 100.5\) (which would now be associated with a severity of 0.89). Or if one wanted to keep the same inference (because that inference is required by the scientific theory or some background knowledge) at the same severity then one would need to observe a far lower \emph{p} value before this could occur\footnote{This final suggestion can take the form of calibrating ones \(\alpha\) level with reference to the sample size and the effect of (scientific) interest. Typically, however, researchers tend to use a fixed \(\alpha\) regardless of context, although recently some have begun to suggest that a single \(\alpha\) level may not be appropriate for all contexts \citep{JYA}.}.

Severity assessments also allow one to draw conclusions about non-significant tests. For instance, when one \emph{fails to reject} \(H_{0}\), it is possible to ask what specific inferences are warranted on the basis of the observed result. Once again using the IQ testing example above, but with a non-significant observation (\(\bar{x} = 102\), n = 100, \(\sigma = 15\)), one can ask what inferences about \(\mu\) are warranted. For example, one might ask whether an inference that \(\mu_{1} < 105\) is warranted or whether the inference that \(\mu_{1} < 103\) is warranted. The severity values associated with each of these inferences (and the observed result) are 0.98 and 0.75, respectively. Therefore, one would have good grounds for inferring that the discrepancy from the null is less than 5, but not good grounds for inferring that it is less than 3. An illustration of severity curves for non-significant observations is shown in Figure 2 (C).

The two examples outlined above are both cases which involve inferences from a single test. But as \citet[p.~S203]{Mayo:1996xx} notes, a ``procedure of inquiry\ldots{} may include several tests taken together''. The use of multiple tests to probe hypotheses with respect to data may be particularly useful in the case where one has failed to reject the null hypothesis. While it is usual to think of significance testing in terms of a null of no effect and an alternative as departures from this null, any value can be designated the null. For example, one might want to test the null hypotheses \(H_{0}:\mu \leq B\) and \(H_{0}:\mu \geq A\) (where usually \(B = - A\)) as a way to examine whether the data support inferences that \(\mu\) lies \emph{within} specified bounds \citep[what can be termed \emph{practical equivalence}, see][]{Phillips:1990ee, Lakens:equi}. This procedure can supplement, or be used as an alternative, to severity interpretations so that one can determine precisely what inferences are warranted on the basis of the data. A consequence of this is that Frequentist inference need not come down to a simple binary (for example, \emph{reject} \(H_{0}\), \emph{fail to reject} \(H_{0}\)/\emph{accept} \(H_{1}\)). Instead, a set of data might lead a researcher to form a far wider range of conclusions. These may include (but are not limited to) inferring: some deviation is present but it is not of sufficient magnitude to support the theory; there are no grounds for inferring that a deviation is present, but neither are there good grounds for inferring any effect lies only within a narrowly circumscribed region around the null; and, there are good grounds for inferring the presence of a deviation from the null and that the deviation is of sufficient magnitude to support a theory.

We will return to Frequentist inference later. For now, one important point to note is that this kind of Frequentist inference is piecemeal. Claims that are more \emph{severely} tested are given more weight than those claims that are not severely tested. Importantly, severe testing might require more than one statistical test---for example, to test assumptions or to break down a hypothesis into multiple piecemeal statistical hypotheses. The severity principle also encourages replication because having to pass multiple tests is a more severe requirement. Activities such as \emph{p}-hacking, optional stopping, or small samples sizes, all directly affect severity assessments by directly changing the error probabilities of the tests. Unfortunately, error statistical thinking has not been common in the psychological literature. However, its value is now starting to be recognised by some \citep[e.g.,][]{Haig:2016kh}, including some working within Bayesian statistics \citep[e.g.,][]{Gelman:2013ka}. Although some of the finer details of the error statistical approach are still to be worked out it may provide a good guide for thinking about how to interpret statistical tests.

\section{An alternative to \emph{p} values}{An alternative to p values}\label{an-alternative-to-p-values}

In the preceding section, we showed that the grounds on which \emph{p} values are granted their epistemic licence are easily violated; however, it has also been argued that \emph{p} values are simply not suitable for scientific inferences because they don't provide the information scientists \emph{really want to know} \citep[e.g., see][]{Nickerson:2000jl, Lindley:2000epl}. On this view, what scientists really want to know is the probability that their hypothesis is true given their data---that is, they want to assign some credence to their hypothesis on the basis of some data they have obtained. Furthermore, \emph{p}-hacking, optional stopping, and similar practices demonstrate the need for procedures that are somehow immune to these behaviours. This alternative, it is claimed, is provided by Bayesian statistics\citep{Dienes:2011cd, Morey:2016gl, Wagenmakers:2007bg}\footnote{In this section, we will use ``Bayesian statistics'' as a shorthand for a suite of approaches that include, but are not limited to, techniques for computing Bayes factors and approaches for estimating the values of unknown parameters. Bayesian statistics should not be taken to mean any procedure that makes use of Bayes Theorem. Bayes Theorem is simply derived from the rules of conditional probabilities. Bayesian statistics, however, is the approach to statistics that aims to produce outputs in the form of degrees of belief and/or degrees of support rather than supporting inferences by controlling certain kinds of errors.}. Bayesian statistics offers a radically different approach to statistical inference, and while largely a niche area in the psychological literature in past decades, events like the replication crisis have sparked renewed interest in these methods.

In offering a solution to what he terms the ``pervasive problem of \emph{p} values'', \citet{Wagenmakers:2007bg} suggests that Bayesian statistics has the desirable attributes for the ideal statistical procedure. These include: 1) that they are dependent only on the observed data, and not the data that might have been collected, 2) that they are immune to the unknown intentions of the researcher, and 3) that they provide a measure of the \emph{strength of evidence} that takes into account both the null and the alternative. Much of the discourse surrounding the switch to Bayesian statistics has focused particularly on the idea that Bayesian statistics may be the solution to problems caused by optional stopping, which have arguably contributed significantly to the replication crisis \citep[e.g.][]{Wagenmakers:2007bg, Rouder:2014jf}. Others, however, have also focused on notions of \emph{evidence} suggesting that the Bayesian conception of strength of evidence is more amenable to scientific reasoning or that it is closer to what researchers intuitively require \citep[e.g.,][]{Morey:2016gl, Szucs:2017it}. It is worth unpacking the claimed advantages of Bayesian statistics in more detail. We will examine the basis of these claims in the sections below.

\subsection{Evidence derived from data alone}\label{evidence-derived-from-data-alone}

In order to unpack the claim that Bayesian inferences are dependent only on the observed data and not data that might have been collected, but wasn't, it is necessary to understand how Frequentist statistics fall into this trap. This Bayesian critique of Frequentist statistics is based on the fact that Frequentist \emph{p} values are calculated from the \emph{sampling distribution}. As outlined earlier, the sampling distribution is a probability distribution of the values of the test statistic under a specified model, such as the null model. It includes all the values that the test statistic might take. And the \emph{p} value is calculated from the tail end probabilities of this distribution---that is, the \emph{p} value expresses: How often would I obtain a value this large \emph{or larger} under this statistical model.

Given this, it is trivial to construct two statistical models (sampling distributions) where the probability of observing a \emph{specific} value of the test statistic is the same, but the chance of observing other values (specifically, larger values) is different. Once a specific value is observed, and a \emph{p} value is calculated, it will be different depending on the probability of obtaining larger values even though the two statistical models \emph{say the same thing} about the observed data. As \citet[p.~385]{Jeffreys:1961tt} put it, the use of ``p implies\ldots{} that a hypothesis that may be true may be rejected because it has not predicted observable results that have not occurred.''

The second desirable property of Bayesian statistics is that, unlike \emph{p} values, Bayesian statistics are not dependent on the unknown \emph{intentions}\footnote{The word \emph{intentions} is often used in the literature. However, it is not the researcher's \emph{intentions} that have an influence on the interpretations of \emph{p} values. Rather, it is researchers' \emph{behaviour} that influences the interpretation of \emph{p} values.} of the researcher. Consider again the case of Alice in the description of the uniformity assumption of the \emph{p} value distribution. Alice collected data from 10 participants, did a significance test and found \emph{p} \textgreater{} .05, added another 10 participants, re-running the test after every participant and then eventually found \emph{p} \textless{} .05. Contrast this with Ashanti, who obtained a sample of 20 participants, ran a significance test and found \emph{p} \textless{} .05. The Frequentist would say that Alice and Ashanti cannot draw the same inferences on the basis of their data, because the severity assessment of Alice and Ashanti's inferences would differ. As \citet{Wagenmakers:2007bg} states, examples like this ``forcefully {[}demonstrate{]} that \emph{within the context of NHST} {[}null hypothesis significance testing{]} it is crucial to take the sampling plan of the researcher into account'' (p.~786). Furthermore, he goes on to state that within the context of Bayesian statistics the feeling people have that ``optional stopping'' amounts to ``cheating'' and that no statistical procedure is immune to this is ``contradicted by a mathematical analysis''. The claim here is that Bayesian statistics are immune to optional stopping and that collecting more data until the patterns are clear is warranted if researchers are using Bayesian statistics.

\subsection{Bayesian statistics and a measure of strength of evidence}\label{bayesian-statistics-and-a-measure-of-strength-of-evidence}

These first two properties of Bayesian statistics, of the immunity to intentions, and of being dependent only on the collected data and not any other data, are derived from what is called the \emph{Likelihood Principle}. The concept of the likelihood allows us to understand the third property of Bayesian statistics, namely that they give us a measure of the strength of evidence. To see this, it is important to know what is meant by \emph{evidence}. As stated in a recent primer for psychological scientists, ``The Likelihood Principle states that the likelihood function contains all of the information relevant to the evaluation of \emph{statistical evidence}. Other facets of the data that do not factor into the likelihood function (e.g., the cost of collecting each observation or the stopping rule used when collecting the data) are irrelevant to the evaluation of \emph{the strength of the statistical evidence}'' \citep[p.~2, our emphasis]{Etz:2017pp}. The intuition here is obvious, if you want to know whether some data supports model \(A\) or model \(B\), all you need to know is whether the data are more likely under model \(A\) or model \(B\). On this view, the strength of evidence is just in the ratio of the likelihoods. If the observed data are three times more likely under model \(A\) than model \(B\), then this can be read as a \emph{measure} of the strength of evidence. Furthermore, if model \(A\) is the null model, then we can say something about the evidential support for this null.

A measure of the strength of evidence is meant to have an additional benefit for the Bayesian. We can weigh our evidence according to some background \emph{pre-data} beliefs we have (e.g., that Model \(A\) is very unlikely to be true) and then use the data to update our beliefs. In Bayesian hypothesis testing, this updating factor is called a \emph{Bayes factor}. Numerically, the Bayes factor can be interpreted as an odds ratio, and it is calculated as the ratio of two \emph{marginal likelihoods} where the marginal likelihood is comprised of a model of the data and some predictions about likely parameter values (sometimes referred to as a \emph{prior} \citep[e.g.,][]{Rouder:2009ij} or a \emph{model of the hypothesis} \citep[e.g.,][]{Dienes:2017bp}). \citet{Rouder:2009ij} give the marginal likelihood for hypothesis \(H\) as:

\[M_H=\int_{\theta\in\Theta_{H}}f_{H}(\theta;\mathbf{y})p_H(\theta)d\theta,\]

where \(\Theta_{H}\) represents the parameter space under the hypothesis \(H\), \(f_{H}\) represents the probability density function of the data under the hypothesis \(H\), and \(p_{H}\) represents the prior distribution of the parameter values expected by that hypothesis. The important point to note here is that calculating a Bayes factor requires the analyst to stipulate some prior probability function for the parameter that they wish to draw inferences about under each of the models they are comparing.

It is worth stepping through this in more detail to understand how this calculation works. To do so, we will consider the task of trying to determine whether a coin is fair (this example, makes use of discrete rather than continuous probability distributions and therefore the integral can be replaced by a sum). For this example, one might define the null hypothesis as \(H_0: \theta = 0.5\), or that the probability of obtaining heads is 50\%. In order to calculate a \emph{Bayes factor}, one needs another hypothesis. We might define this hypothesis as the probability of obtaining heads being some other fixed value---for example, \(H_1: \theta = 0.7\), or that the probability of obtaining heads is 70\%. If we were to further consider \(H_0\) and \(H_1\) equally plausible, our Bayes factor value would simply be the likelihood ratio of these two hypotheses. For example, given a set of data such as the observation of 2 heads out of 10 flips we could conclude that this observation is 30.38 times more probable under the hypothesis that \(\theta=0.5\) than the hypothesis that \(\theta=0.7\).

However, we are ordinarily not interested in a single parameter value but are instead concerned with models in which the parameter may take one of several different values. In our coin flipping example, this might mean comparing \(H_0:\theta = 0.5\) and an alternative hypothesis \(H_1\) composed of 11 point hypotheses (\(H:\theta=0\), \(H:\theta=0.1\), \(H:\theta=0.2\), \(\ldots\) \(H:\theta=1\)) spanning the entire range of values that \(\theta\) might take. To calculate the \emph{Bayes factor}, we first calculate the likelihood ratio of the data under \(H_0\) to each of the 11 point hypotheses of \(H_1\). The \emph{Bayes factor} is then computed as the \emph{weighted sum} of these 11 values, where the weights are determined by a \emph{prior} assigned to each of the 11 point hypotheses that make up \(H_1\). The prior makes predictions about what parameter values (bias values in our example) are expected under \(H_1\)\footnote{In this context, \emph{prior} refers to the weights we assign to each of the likelihood ratios for each of the possible parameter values. The term \emph{prior} (sometimes \emph{prior odds}) is also used to refer to our predata beliefs about how likely we think it is that \(H_0\) or \(H_1\) is true. This second type of \emph{prior} doesn't factor into the calculation of the Bayes factor but, as noted above, can be used in conjunction with a Bayes factor to determine our post data beliefs. Consequently, if we think that biased coins are infinitesimally rare then even obtaining a large Bayes factor in favour of \(H_1\) would not lead us to conclude that we have encountered a biased coin.}. If for example, we were to consider each possible value of \(\theta\) to be equally likely under our biased coin model, then we would weigh each likelihood ratio equally. Because the prior is a probability distribution, the weights should sum to one, which means that each likelihood ratio would have a weight of 1/11. For our example of observing 2 heads in 10 flips this would correspond to a Bayes factor of 2.07 in favour of \(H_1\).

This uniform prior is just one example of a prior one might choose. One might decide that the uniform prior is not very realistic and instead decide to employ a non-uniform prior. In our coin flipping example, we might use a prior that places more weight on values further from 0.5 than values closer to 0.5 if we believe trick coins are likely to be heavily biased (for example, a beta prior such as \(\theta\sim\mathrm{Beta}(0.9,0.9)\)). We might use a prior that represents our belief that trick coins will be heavily biased towards coming up heads (for example, a beta prior such as \(\theta\sim\mathrm{Beta}(5,1)\)). Or we might believe that trick coins are unlikely to be heavily biased and instead use a prior that places most of its weight at values near 0.5 (for example, a beta prior such as \(\theta\sim\mathrm{Beta}(10,10)\)). In each of these cases the Bayes factor will be different: We would obtain values of 0.5, 8.78, and 0.66 in favour of \(H_0\) for each one of these three models or priors. In these examples, we have chosen to use the prior to quantify our beliefs about outcomes that are likely to occur when coins are unfair (that is, they are our models of what unfair coins are like). As \citet{Dienes:2017bp} points out, this requires the analyst to specify the predictions of the models being compared and thus the Bayes factor can be interpreted as the relative predictive accuracy of the two models. That the models have to make predictions about what data is likely to be observed has the added benefit that models that are overly vague are penalised. This can simply be illustrated by modifying the width of a prior so that a model predicts an increasingly wide range of implausible values. An example of this (using the default Bayesian \emph{t}-test discussed below) is shown in Figure 3.

\begin{figure}
\centering
\includegraphics[width=.45\textwidth]{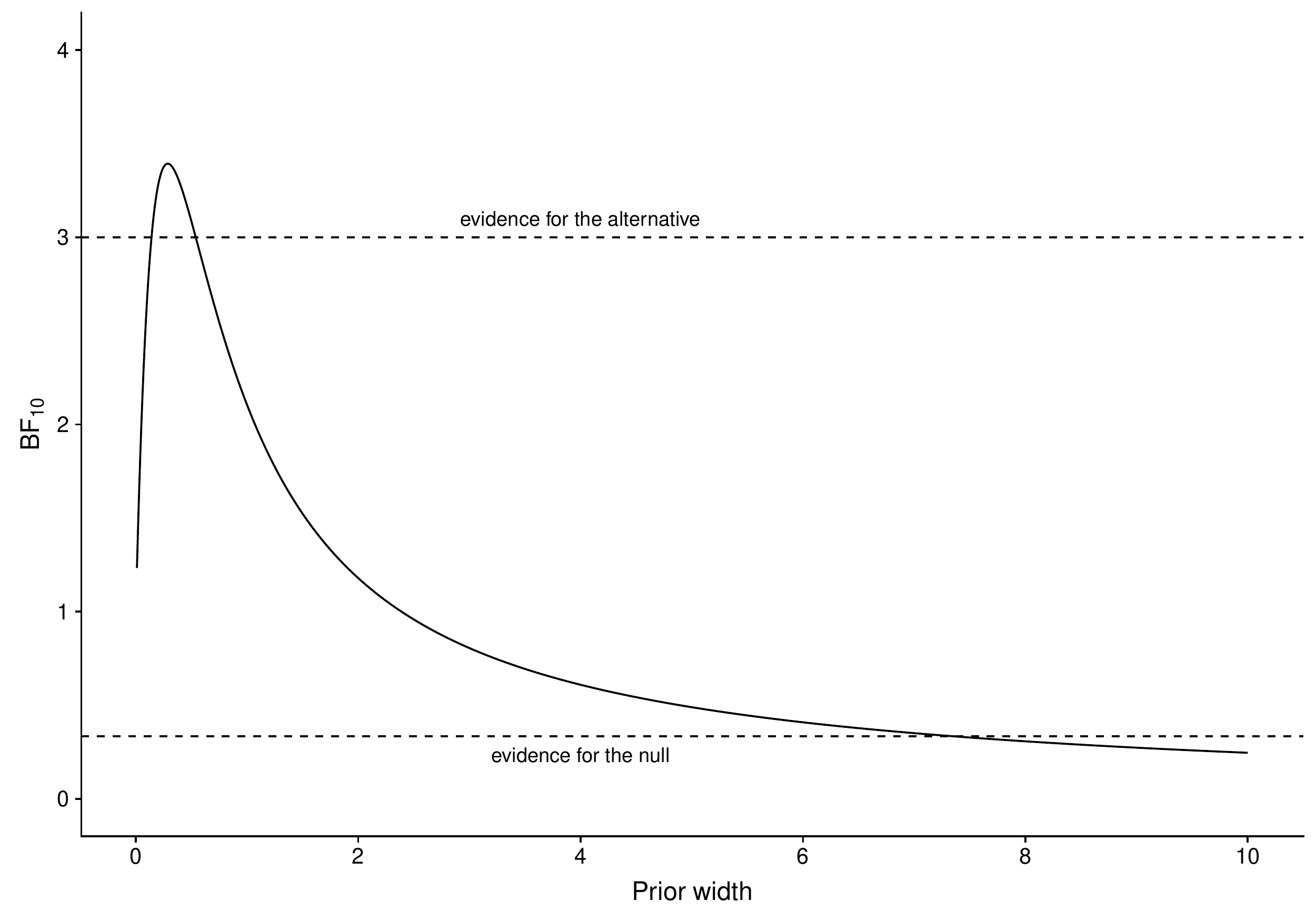}
\caption{Bayes factor values as a function of prior width.}
\end{figure}

There are two broad schools of thought about how one should go about specifying these model predictions. \emph{Subjective} Bayes approaches seek to employ priors that reflect the analyst's prior beliefs about likely parameter values \citep{Rouder:2009ij, Dienes:2017bp, Berger:2006pri, Dienes:2014non, Gronau:2018}, as we have done with our coin flipping example. The \emph{objective} Bayesian approach, on the other hand, seeks priors that are minimally informative\footnote{Minimally informative (or non-informative) is used here in the technical sense to refer to, for example, Jeffreys' prior, not in the colloquial sense of being vague. A \emph{subjective prior} might be non-informative in the colloquial sense without being non-informative in the technical sense.}. Often priors are sought that are appropriate in as wide a range of scenarios as possible or priors that have good frequentist properties \citep{Berger:2006pri}. One such example is the JZS prior on the effect size parameter, which is found in the default Bayesian \emph{t}-test \citep{Rouder:2009ij}.

The fact that inference from Bayes factors depends on model specifications is not inherently problematic. As our coin flipping example shows, deciding whether a coin is fair or not is dependent on what we think it means for a coin to be unfair. That is, our inferences are specific to the models being compared. However, some difficulties can arise when it comes to specifying the models that are to be compared by the analysis. It is worth examining how disagreements about model specifications can give rise to different inferences by examining a few examples taken from \citet{Dienes:2017bp}. These examples will also be instructive because they were selected to highlight some of the putative benefits of the Bayesian approach over the Frequentist approach.

The first example reported by \citet{Dienes:2017bp} is of an experiment where participants in two conditions were required to make judgements about the brightness of a light. \citet{Dienes:2017bp} report the results from both the original finding and a subsequent replication attempt. In the original paper, the authors report a difference between the two conditions in brightness judgement of 13.3 Watts, and a corresponding statistically significant \emph{t}-test (\emph{t}(72) = 2.7, \emph{p} = .009, cohen's d = 0.64). For the replication attempt the sample size was increased such that if the true effect was of the same magnitude as the original finding the replication attempt would produce a statistically significant result approximately 9 times out of 10---that is, the statistical power would be 0.9. The replication attempt, however, failed to produce a statistically significant result(\emph{t}(104) = 0.162, \emph{p} = 0.872, cohen's d = 0.03), and a raw effect of approximately 5.47 Watts was observed. What is one to make of this failed replication attempt?

\citet{Dienes:2017bp} state in the case of the second experiment that ``{[}b{]}y the canons of classic hypothesis testing {[}that is, frequentist methods{]} one should accept the null hypothesis.'' As noted earlier in our discussion of Frequentist inference, a non-significant result does not warrant the inference \emph{accept} \(H_{0}\), at least not from a principled perspective. However, setting this aside, for now, we can ask what the Bayesian should conclude. According to the analysis presented by \citet{Dienes:2017bp}, the original finding, which reported a raw effect of 13.3 Watts, should inform the analyst's model of \(H_{1}\). The resulting Bayes factor computed on the basis of this model after observing the new data (the raw difference of 5.47 Watts) is approximately 0.97. That is, the Bayes factor value indicates that the new data provide roughly equal support for the null and the alternative and the conclusion should be that the results are inconclusive. \citet{Dienes:2017bp} may be justified in this specification of an informed prior; however, one might, either through a desire for ``objectivity'' or through a desire to compare one's inference to some reference, instead choose to use a non-informative prior. The JZS prior, employed in the default Bayesian \emph{t}-test \citep{Rouder:2009ij}, is one such example. Re-running the analysis employing this new model specification for the alternative hypothesis now instead results in a Bayes factor of 0.21---that is, the null is now preferred by a factor of nearly 5 to 1. Interestingly, this is just the same inference as the \emph{heuristic} interpretation of the \emph{p} value.

It is important to note, however, that the fact that the two Bayesian analyses give different results is not a problem, at least not from a Bayesian perspective. The analysis is simply providing a measure of the strength of evidence for one model relative to another model. A problem only arises when one seeks to interpret the Bayes factor as an indication of \emph{``an effect'' being present} versus \emph{``an effect'' being absent}. However, it is also worth noting that with default priors (that is, the JZS prior), the model being compared is not really a model of the theory in the same sense as Dienes and Mclatchie's \citeyearpar{Dienes:2017bp} model is, which somewhat breaks the connection between the statistical hypothesis and the scientific hypothesis. However, since any change in statistical practice is likely to depend on ease-of-use (both in terms of conceptual understanding and the availability of, for example, software tools) it seems likely that default priors may be the dominant type of model specification in use, at least in the short term. And therefore, it is necessary that the appropriate caveats are observed when drawing inferences on the basis of these procedures.

Just as Bayesian inference is relative to specific models, it is also important to reiterate that Frequentist inferences should be relative to specific alternative hypotheses that are assessed against actual observed results. This more sophisticated frequentist analysis would actually draw conclusions more similar to the inferences drawn by \citet{Dienes:2017bp}. For example, the Frequentist might want to use severity assessments to assess various hypotheses with respect to the observed result. If this was done, the inference, like the Bayesian inference would be similarly inconclusive. Inferences about only very small discrepancies being present are not tested with severity (that is, inferences that accord more with the null hypothesis would not be supported). The only inferences that would pass with severity are those that entertain the possibility of a wide range of discrepancies---from negligible to very large---being present (that is, an inconclusive result). Furthermore, a more sophisticated Frequentist might also choose to perform multiple statistical tests to test this one scientific hypothesis, and to build up evidence in a piecemeal manner. One way to do this would be to perform two one-sided tests against the twin null hypotheses of, for example, \(H_{0}:\mu > - 10\) Watts and \(H_{0}:\mu < 10\) Watts. This would allow the analysts to draw inferences about practical equivalence within the range of, for example, -10 to +10 Watts. The results of such an equivalence test would be non-significant suggesting that the null hypotheses cannot be rejected and again suggesting that the result is inconclusive (\emph{t}(104) = -0.13, \emph{p} = 0.45).

It is an interesting exercise to apply severity reasoning to the other examples presented by \citet{Dienes:2017bp}. For instance, \citet{Dienes:2017bp} shows that a Bayesian analysis can be used to show that a non-significant effect from an experiment with low \emph{a priori} power need not be viewed as evidentially weak. However, severity assessments for non-significant results do not rely on \emph{pre-experiment} power (that is, a power calculation performed before the data is known), as a naïve Frequentist might, but rather assess hypotheses with respect to the data \emph{actually obtained}. For this example, it is possible to probe various hypotheses to see which pass with severity. Applying this reasoning to the same example as \citet{Dienes:2017bp} would result in concluding that the data are consistent with the presence of a negligible to very small effect, but not consistent with a large effect. Or one might use multiple tests, taken together, such as in an equivalence test procedure, and find that one has good grounds to infer that any deviations from the null fall within the bounds of practical equivalence\footnote{In fact, running such an equivalence test on the data presented in their example does result in one rejecting the null hypothesis of an effect larger than practical equivalence (±1\% difference between groups in the number of questions answered correctly) being present (\emph{t}(99) = 1.72, \emph{p} = 0.04).}. Furthermore, severity assessments of a \emph{just} significant effects in a large study would lead one to conclude that there are not good grounds for inferring that anything but a negligible effect is present just as a significant (Frequentist) effect in a large study would lead to a Bayes factor that strongly favours the null model over the alternative model.

\section{Two approaches to inference, evidence, and error}\label{two-approaches-to-inference-evidence-and-error}

We have outlined a view of inference offered from the Frequentist, error-statistical, perspective in the form of the severity principle: One can only make claims about hypotheses to the extent that they have passed severe tests. And we have outlined a view of inference offered from the Bayesian perspective: One can make claims about hypotheses to the extent that the data support that hypothesis relative to alternatives. These two approaches are often pitched as rivals because it is argued that they can warrant different inferences when presented with the same data, as the examples presented by \citet{Dienes:2017bp} are meant to show. However, as our discussion of \citet{Dienes:2017bp} shows, this is not clearly the case. What these examples more clearly demonstrate is that the exact nature of the \emph{question} being asked by Dienes and Mclatchie's \citeyearpar{Dienes:2017bp} Bayesian analysis and the naïve frequentist analyses they present are different. With different questions one need not be surprised by different answers. The same applies to asking two different Bayesian questions (one using a default prior and one using an informed prior)---a different question results in a different answer. Consequently, when \citet{Dienes:2017bp} point out pitfalls of significance testing they are in fact pointing out pitfalls associated with a naïve approach. A more sophisticated use of Frequentist inference allows one to avoid many of the common pitfalls usually associated with significance testing and it is not necessary to adopt Bayesian methods if all one wants to do is avoid these misinterpretations.

There are, however, situations where Bayesian and Frequentist methods are said to warrant different inferences that are a consequence of the \emph{process} that allows each type of inference to be justified. Consider, for example, the claim of \citet{Wagenmakers:2007bg} that the feeling that optional stopping is cheating is contradicted by a mathematical analysis. From an error statistical perspective any claims made as a result of optional stopping are not warranted (making those claims \emph{is} cheating) because the claims have not been severely tested (the probability of detecting an error would be very low so not detecting an error is unimpressive). The same applies for data-dredging and a range of other behaviours. For the Bayesian, however, all that matters in assessing the \emph{strength of evidence} is the ratio of the likelihoods. The Bayesian can be seen as regarding \emph{data} as primary while the Frequentist can be seen as regarding the \emph{process} as primary. As noted by \citet{Haig:2016kh}, this is a difference between Frequentists (specifically, of the error-statistical variety) favouring local or context-dependent accounts of statistical inferences with Bayesians' favouring broad general or global accounts of statistical inference.

The important question, however, is how does each approach fair as a system of \emph{scientific inference}? The primary difference between the two can be seen as coming down to error control. Frequentists, like Mayo \citep{Mayo:1996xx, Mayo:2011wm, Mayo:2006iu} insist that any system of inference must be so designed so that we are not lead astray by the data. Consider the case of collecting observations and then drawing inferences on the basis of these. It might be reasonable to ask whether those observations reflect some truth or whether they are possibly misleading. Bayesian statistics, however, does not care about error probabilities in assessing the strength of evidence. The strength of evidence (derived from the Likelihood Principle) is simply construed as the degree to which the data support one hypothesis over the other with no reference to how often the evidence might be \emph{misleading}. This is in distinction to Frequentist approaches that fix at an upper-bound how often inferences will be in error. This highlights what \citet{Birnham:evide} called the ``anomalous'' nature of statistical evidence. \citet{Gandenberger:2015cz}, similarly, cautions against using the Likelihood Principle to argue against Frequentist statistics, particularly the error statistical view. Whether the Likelihood Principle is true or not, is simply not relevant for this system of inference and, therefore, Frequentist violations of the likelihood principle are of no consequence \citep{Gandenberger:2015cz}. Similarly, ignoring error probabilities is of no consequence within the Bayesian system of inference \citep{Haig:2016kh}. \citet{Gandenberger:2015cz} states that the likelihood principle only applies if one wants to use methods that track ``evidential meaning'', but he goes on to state that while ``tracking evidential meaning is intuitively desirable\ldots{} {[}it may be{]} less important than securing one or more of {[}the{]} putative virtues'' of Frequentist methods. These virtues, such as the ability to control error probabilities and the ability to \emph{objectively} track truth (in, for example, the absence of priors), may be virtues that one wishes to retain.

The Bayesian view that the evidential import of the data is only reflected through the likelihoods is also more nuanced than is often recognised. Specifically, the adherence to the Likelihood Principle implies an immunity to stopping rules; however, this immunity must be qualified. There are many instances when the stopping rule may influence the inferences that the Bayesian wants to draw from the data obtained in an experiment. In these situations, the stopping rule is described as \emph{informative}. Stopping rules are said to be \emph{informative} if, for example, they provide information about a relevant unknown parameter that is not contained within the data itself. For example, when trying to estimate some parameter, \(\theta\), if the stopping rule is dependent on \(\theta\) in some way other than through the data, such as by making some stopping rule more likely if \(\theta = X\) and another stopping rule more likely if \(\theta = Y\), then the stopping rule carries information about \(\theta\) that is not in the data itself. To adapt an example from \citet{Edwards:1963gk}: If you are trying to count the number of lions at a watering hole, then the fact that you had to stop counting because you were chased away by all the lions should factor into any of your inferences about the number of lions. \citet{Roberts:1967hl} presents some more formal examples and suggests that in these cases it is right and proper to take this parameter dependence into account in the likelihood function.

Information about the stopping rule can also enter into a Bayesian inference through the prior more directly when objective priors are used. Consider the example of flipping a coin multiple times and after each flip recording whether it landed on \emph{heads} or \emph{tails}. Once the data is obtained, one might want to make an inference about the probability of obtaining heads. As pointed out by \citet{Wagenmakers:2007bg}, for a Frequentist to draw inferences about the observed data they would need to have information about how the data was collected---that is, the stopping rule. Specifically, it would be necessary to know whether, for example, the data were collected until a fixed number of trials were completed or until a fixed number of heads were recorded. The two sampling rules can lead to identical observed data, but since the two sampling rules have something different to say about \emph{possible} data that \emph{could} occur under the null hypothesis, this information must enter into the Frequentist analysis. \citet{Etz:2017pp} also makes use of this example, not to show the flaw in Frequentist inference (which is what \citet{Wagenmakers:2007bg} deploys the example for), but to show how a Bayesian can make use of prior information when computing the posterior probability of obtaining heads. In his example, \citet{Etz:2017pp} shows how one can combine some prior beliefs (for example, the belief that the probability of obtaining heads is likely to be between 0.30 and 0.70) to obtain a posterior distribution of values for obtaining heads. In Etz's \citeyearpar{Etz:2017pp} example, his prior quantifies his \emph{pre-data beliefs}, and his posterior quantifies his \emph{post-data beliefs} that have been updated in light of the data. However, how is one to perform the Bayesian analysis if one has no pre-data beliefs or no strong grounds for holding a particular set of pre-data beliefs?

As mentioned earlier, the use of objective priors is meant to circumvent the problems of specifying these subjective priors. The solution, therefore, is just to make use of one of the minimally informative objective priors. \citet{Box:1973pri} provide just such a set of non-informative priors derived from Jeffreys' rule; however, the exact prior that is appropriate turns out to be dependent on the sampling rule. That this, the ``objective'' Bayesian inference about the parameter from a set of data turns out to be different depending on how the data were collected. As noted by \citet{Hill:1974rev} and \citet{Berger:2006pri}, this amounts to a violation of the Likelihood Principle. In Wagenmakers's \citeyearpar{Wagenmakers:2007bg} terms, it would result in a Bayesian analysis that is dependent on the \emph{unknown intentions} of the researcher. \citet[p 46]{Box:1973pri} note that they find the observation that a difference in sampling rules leads to different inferences ``much less surprising than the claim that they ought to agree.'' Indeed the requirement that one adheres to the Likelihood Principle in drawing inferences is not universally accepted even among Bayesian's. For example, \citet{Gelman:2013ka} encourage a kind of data-dependent model validation that might similarly violate the Likelihood Principle when the entire inference process is viewed as a whole. Furthermore, \citet[p 198]{Gelman:BDA} state, ``\,`the observed data' should include information on how the observed values arose''. That is, good Bayesian inference should be based on all the available information that may be relevant to that inference. However, the assessment of evidence, once data and models are in hand can still be done in a manner that respects the Likelihood Principle.

In addition to cases where \emph{informative} stopping rules are used, cases may also arise where stopping rules that are ostensibly \emph{uninformative} from one perspective might be informative from another perspective. These kinds of situations are likely to arise more often than is often recognised. \citet{Gandenberger:2017vt} outlines such a situation. Consider two researchers, Beth employs the stopping rule: collect data until the likelihood ratio favours \(H_{1}\) over \(H_{0}\) by some amount. Brooke employs the stopping rule: collect data until reaching some fixed \emph{n}. The stopping rule employed by Beth is technically \emph{uninformative} because the stopping rule is only dependent on the data observed and is not dependent on other information about the parameter of interest not contained in the data. If it happens to be the case that Beth and Brooke obtain identical data then the Bayesian analysis states that Beth and Brooke are entitled to identical inferences.

However, consider a third party, Karen, who is going to make decisions on the basis of the data. For Karen, it might not be that easy to discount the stopping rule. For example, if she suspects that Beth might choose her stopping rule on the basis of a pilot experiment that showed evidence in favour of \(H_{0}\) then the stopping rule contains information that is of some epistemic value to Karen. This situation, where there is a separation between inference-maker and data collector, is not uncommon in science. Other researchers who will make inferences on the basis of published research, journal editors, reviewers, or other end users of research may consider a stopping rule informative even when the researcher themselves does not.

Other instances might also exist where a Bayesian might want to consider stopping rules. One such example is suggested by \citet{Berger:1988cz}. They suggest that if somebody is employing a stopping rule with the aim of making some parameter estimate exclude a certain value then an analyst might want to take account of this. For example, \citet{Berger:1988cz} suggest that if a Bayesian analyst thinks that a stopping rule is being used because the experimenter has some belief about the parameter (for example, that the estimate should exclude zero), then adjustments should be made so that the posterior reflects this. These adjustments, however, should not be made to the likelihood---that is, they should not affect the \emph{strength of evidence}---but should instead be made to the prior so that some non-zero probability is placed on the value that the experimenter might be trying to exclude. This approach, however, has not been without criticism. Specifically, the practice of making adjustments to priors because an analyst might \emph{think} that an experimenter \emph{thinks} something about a parameter runs a severe risk of appearing ad hoc. This is especially the case given that much of the Bayesian criticism of Frequentist statistics is based on the claim that unknown \emph{intentions} should not influence inferences. The Frequentist response is much more satisfactory. After all, the Frequentist can point to specific problematic \emph{behaviour} that justifies their rule; however, \citet{Berger:1988cz} appear to suggest that the Bayesian really must care about the mental states of the data collector.

The upshot of examples like this is that far from immunity to stopping rules, the conditions under which stopping rules are informative can be poorly defined. Furthermore, the responses to these situations can be tricky to implement. The fact remains that many of the cases where Frequentists are worried about stopping rules may be the very same cases where stopping rules \emph{should} worry a Bayesian too.

\subsection{What do we really want to know?}\label{what-do-we-really-want-to-know}

What should we make of examples where stopping rules appear to influence the epistemic value of the data? One solution is to ask ourselves what we really need for scientific inference. For example, \citet{Gandenberger:2015cz} recognises that it is reasonable to care about error probabilities despite them having no influence on \emph{evidence}. And \citet[p 286]{Dienes:2011cd} suggests that ``{[}u{]}ltimately, the issue is about what is more important to us: using a procedure with known long term error rates or knowing the degree of support for our theory.'' There are several legitimate reasons for deciding that \emph{both} are important.

The reasons for wanting to know both is that the two kinds of inferences figure differently in scientific reasoning. Caring about error rates is important because one can learn from the absence of error, but only if there is a good chance of detecting an error if an error exists \citep[e.g.,][]{Mayo:1996xx}. When one collects observations it may be less important to know whether or not a particular observation is better predicted by theory A or theory B. Instead, it may be better to know whether inferences about the presence or absence of error are well justified, which is what can be gained from the severity principle. For instance, if we wish to conclude that an observation justifies a conclusion of some deviation from a particular model then whether we have good grounds for this inference can be determined with reference to the severity principle. Similarly, if we wish to conclude that we have good grounds for inferring that there is no deviation (within a particular range), then the severity principle can help here too. And all this can be done without needing to know whether and to what extent that deviation is predicted by two theories.

However, if one has good grounds for making one's models and good grounds for making predictions, then it seems reasonable to care about whether the evidence supports one model over its rival. With some observations in hand, along with some explanations or models, a Bayesian analysis allows us to judge which is the best explanation. \citet{Haig:2016kh} similarly echoes this view that both forms of inference are necessary by calling for \emph{pragmatic pluralism}. However, for this to work it is important to understand the strengths and weaknesses of each approach, the inferences each approach warrants, and when each approach should be deployed. This, however, is a different kind of argument than that which is ordinarily made by those advocating statistical reform \citep{Wagenmakers:2007bg, Dienes:2017bp}. The usual strategy here is to argue that Bayesian statistics should be adopted because they lead to more reasonable, more correct, or more intuitive inferences from data relative to Frequentist inference. As we have pointed out in Section 3.2, in our discussion of \citet{Dienes:2017bp}, the Frequentist inference and the Bayesian inference can often be similar on a gross level (the data are inconclusive, the data support an alternative hypothesis, the data do not support an alternative hypothesis) and, therefore, arguing that statistical reform is necessary because macro level inferences are different may not work as a strategy. A better strategy, we believe, is to argue that statistical reform is necessary because it is necessary to have the right tool for the right job in a complete system of \emph{scientific inference}.

Tests of statistical significance find their strength where reasonable priors are difficult to obtain and when theories may not make any strong quantitative predictions. For example, when researchers simply want to know whether they can reliably measure some phenomenon (of a specific magnitude or range) then significance testing might play a role. \citep[Significance tests play an analogous role in physics, see][]{vanDyk:2014kl}. In these contexts, however, it is important that researchers at least have some sense of the magnitude of the effects that they wish to observe so that analyses can be adequately powered. Furthermore, they might be useful in exploratory contexts. This kind of exploratory research is importantly different to data dredging---that is, rather than testing numerous statistical hypotheses, finding significance, and then claiming support for a substantive hypothesis, this kind of exploratory research involves the systematic collection of observations. Importantly, the systematic collection of observations will involve piecemeal accumulation of evidence, coupled with repeated tests and follow-ups to ensure severity. In the psychological sciences, one such context might be neuro-imaging\footnote{This is just used as a hypothetical example. Whether this works in practice depends crucially on the ability to control error rates. While controlling error rates is \emph{in theory} possible, in practise, this has proved more difficult \citep[e.g.,][]{Eklund:2016bs}.} where a researcher simply wants to know whether some response can be reliably measured with the aim of later building a theory from these observations \citep[see][]{Colling:2010tt}. This is essentially a signal detection task and it does not require that one specify a model of what one expects to find. Instead, the minimal requirement is a model of the noise, and the presence of signals can be inferred from departures from noise. Importantly, theories developed in this way could then be tested by different means. If the theory takes the form of a quantitative model or, better yet, multiple competing plausible models then a switch to Bayesian statistics would be justified.

Bayesian statistics thrives in situations involving model comparison, parameter estimation, or when one actually wishes to assign credences, beliefs, or measure the degree of support for hypotheses. Significance testing has no formal framework for belief accumulation. However, to fully exploit these strengths psychological scientists would not only need to change the way they do statistics but also change the way they do theory. This would involve an increased emphasis on explanation by developing \emph{quantitative} mechanisms \citep[see][]{Kaplan:2011ke, Colling:2014kd}. Unfortunately, the naïve application of significance tests does not encourage the development of mechanistic theories that make quantitative predictions. Rather, the focus on simple dichotomous reject/do not reject thinking can, and has, lead researchers to often be satisfied with detecting \emph{any} effect rather than \emph{specific} effects.

Importantly, the debates around statistical reform and the replication crisis highlight a deeper concern. Rather than \emph{merely} a statistical issue, the replication crisis highlights the stark disconnect between those inferences that are \emph{warranted} and \emph{justified} and those inferences that scientists actually make, both with respect to their own work and with respect to the work of others. \citet{Haig:2016kh} and \citet{Szucs:2017it} raise similar concerns. Rather than offloading inferences onto sets of numbers produced by statistical procedures, researchers, and particularly students, need to have a greater understanding of how to construct appropriate explanatory theories and how to differentiate substantive and statistical hypotheses. Additionally, it is also important that researchers are able to identify contexts in which hypothesis tests (whether Bayesian or Frequentist) are appropriate and contexts in which parameter estimates are more appropriate---that is, when to \emph{test} hypotheses and when to \emph{measure} phenomena.

\section{Conclusions}\label{conclusions}

We do not think that the solution to the replication crisis lies in statistical reform per se. While there are undoubtedly problems with how people justify scientific inferences on the basis of statistical significance tests, these problems may lie less with the tests themselves than with the inferential systems people employ. And we have attempted to demonstrate how good inferences on the basis of statistical significance tests may be justified. We have also examined the Bayesian alternative to statistical significance tests and explored some of the benefits of the Bayesian approach. The argument for Bayesian statistics is often framed in terms of the macro level inferences that they permit and in terms of the perceived shortcomings of Frequentist statistics. However, we have argued that well-justified Frequentist inferences can often lead to the same gross conclusions. Rather, the key differences lie in their view of evidence and the role error plays in learning about the world. That is, rather than furnishing different inferences, per se, each approach provides a different kind of information that is useful for different aspects of scientific practice. Rather than mere statistical reform, what is needed is for scientists to become better at inference (both Frequentist and Bayesian) and for a better understanding of how to use inferential strategies to justify knowledge.

\section{Acknowledgements}\label{acknowledgements}

We would like to thank two anonymous reviewers for their comments on their manuscript. Their thoughtful comments greatly improved this manuscript.

\bibliographystyle{apa}
\bibliography{cited} 

\end{document}